%% file: acl_latex.tex
\pdfoutput=1
\documentclass[11pt]{article}

\usepackage{acl}
\input{math_commands.tex}

\usepackage{times}
\usepackage{latexsym}
\usepackage[utf8]{inputenc}
\usepackage[T1]{fontenc}
\usepackage{hyperref}
\usepackage{url}
\usepackage{booktabs}
\usepackage{amsfonts}
\usepackage{nicefrac}
\usepackage{microtype}
\usepackage{xcolor}
\usepackage{enumitem}
\usepackage{multirow}
\usepackage{graphicx}
\usepackage{subfigure}
\usepackage{pgfplots}
\pgfplotsset{compat=1.7}
\usepgfplotslibrary{groupplots}
\input{colors}
\input{figures/util}

\usepackage{amsmath}
\usepackage{amssymb}
\usepackage{mathtools}
\usepackage{amsthm}
\usepackage{footmisc}
\usepackage[skins]{tcolorbox}

\newcommand{\commentout}[1]{}

\def\code#1{\texttt{#1}}

\title{Large Language Models are Effective Text Rankers with Pairwise Ranking Prompting}

\author{Zhen Qin, Rolf Jagerman, Kai Hui, Honglei Zhuang, Junru Wu, Le Yan, Jiaming Shen,\\ 
 \textbf{Tianqi Liu, Jialu Liu, Donald Metzler, Xuanhui Wang, Michael Bendersky}\\
 Google Research \\
 \{zhenqin,jagerman,kaihuibj,hlz,junru,lyyanle,jmshen,tianqiliu,jialu,metzler,xuanhui,bemike\}@google.com
 }

\begin{document}
\maketitle
\begin{abstract}
Ranking documents using Large Language Models (LLMs) by directly feeding the query and candidate documents into the prompt is an interesting and practical problem. However, 
researchers have found it difficult to outperform fine-tuned baseline rankers on benchmark datasets.
We analyze pointwise and listwise ranking prompts used by existing methods and argue that off-the-shelf LLMs do not fully understand these challenging ranking formulations. In this paper, we propose to significantly reduce the burden on LLMs by using a new technique called \emph{Pairwise Ranking Prompting} (PRP).
Our results are the first in the literature to achieve state-of-the-art ranking performance on standard benchmarks using moderate-sized open-sourced LLMs. On TREC-DL 2019\&2020, PRP based on the Flan-UL2 model with 20B parameters performs favorably with the previous best approach in the literature, which is based on the blackbox commercial GPT-4 that has 50x (estimated) model size, while outperforming other LLM-based solutions, such as InstructGPT which has 175B parameters, by over 10\% for all ranking metrics. By using the same prompt template on seven BEIR tasks, PRP outperforms supervised baselines and outperforms the blackbox commercial ChatGPT solution by 4.2\% and pointwise LLM-based solutions by more than 10\% on average NDCG@10.
Furthermore, we propose several variants of PRP to improve efficiency and show that it is possible to achieve competitive results even with linear complexity.
\end{abstract}

\section{Introduction}
\label{sec:intro}
Large Language Model (LLMs) such as GPT-3~\citep{brown2020language} and PaLM~\citep{chowdhery2022palm} have demonstrated impressive performance on a wide range of natural language tasks, achieving comparable or better performance when compared with their supervised counterparts that are potentially trained with millions of labeled examples, even in the zero-shot setting~\citep{kojima2022large, agrawal2022large, huang2022language, hou2023large}.

However, there is limited success for the important text ranking problem using off-the-shelf LLMs~\citep{jimmygpt}. Existing results usually significantly underperform well-trained baseline rankers (e.g.,~\citet{monot5, zhuang2022rankt5}). The only exception is a recent approach proposed by~\citet{baidugpt}, which depends on the blackbox commercial GPT-4 system. Besides the technical concerns such as sensitivity to input order (ranking metrics can drop by more than 50\% when the input document order changes), we argue that relying on such blackbox systems is not ideal for academic researchers due to significant cost constraints and access limitations to these systems, though we do acknowledge the value of such explorations in showing the capabilities of LLMs for ranking tasks.

In this work, we first discuss why it is difficult for LLMs to perform ranking tasks with existing methods, specifically, the pointwise and listwise formulations. For pointwise approaches, ranking requires LLMs to output \emph{calibrated} prediction probabilities before sorting, which is known to be very difficult and is not even supported by the generation-only LLM APIs (such as GPT-4). For listwise approaches, even with instructions that look very clear to humans, LLMs can frequently generate conflicting or useless outputs, which happens especially often for moderate-sized LLMs that are used in our experiments. Such observations show that existing popular LLMs do not fully understand ranking tasks, potentially due to the lack of ranking awareness during their pre-training and (instruction) fine-tuning procedures. 

We propose the Pairwise Ranking Prompting (PRP) paradigm, which uses the query and a pair of documents in the prompt for LLMs to perform ranking tasks, with the motivation to significantly reduce the task complexity for LLMs and resolve the calibration issue. PRP is based on simple prompt design and naturally supports both generation and scoring LLMs APIs. We describe several variants of PRP to address efficiency concerns. PRP results are the first in the literature that can achieve state-of-the-art ranking performance by using \textit{moderate-sized, open-sourced} LLMs on standard benchmark datasets. On TREC-DL2020, PRP based on the FLAN-UL2 model with 20B parameters outperforms the previous best approach in the literature, based on the blackbox commercial GPT-4 that has (an estimated) 50X model size, by over 5\% at NDCG@1. On TREC-DL2019, PRP is only inferior to the GPT-4 solution on the NDCG@5 and NDCG@10 metrics, but can outperform existing solutions, such as InstructGPT which has 175B parameters, by over 10\% for nearly all ranking metrics. We also show competitive results using FLAN-T5 models with 3B and 13B parameters, demonstrating the power and generality of PRP. The observations are further validated on seven BEIR datasets covering various domains, where PRP performs competitively with supervised rankers and outperforms other LLM based approaches by a large margin. We further discuss other benefits of PRP, such as being insensitive to input ordering. 

We note that "pairwise" paradigm is in itself a very general and classic idea that impacted a wide range of areas. The novelty of our work lies in the important scenario where the technique is introduced, the adaptations to make it practical, the effectiveness it enables, as well as potential changes and insights it inspires. In summary, the contributions of this paper are three-fold:
\begin{itemize}[noitemsep,topsep=0pt,parsep=0pt,partopsep=0pt]
    \item We for the first time in published literature show pairwise ranking prompting effectiveness for ranking with LLMs. It is able to produce state-of-the-art ranking performance on a wide range of datasets with simple prompting and scoring mechanism.
    \item Our results are based on moderate-sized, open-sourced LLMs, comparing with existing solutions that use blackbox, commercial, and larger models. The finding will facilitate future research in this direction.
    \item We study several efficiency improvements and show promising empirical performance.
\end{itemize}

\section{Difficulties of ranking tasks for LLMs}
\label{sec:analysis}
As discussed in Section~\ref{sec:intro}, to date there is limited evidence showing off-the-shelf LLM-based rankers can outperform fine-tuned smaller rankers. We discuss why this is the case by overviewing and analyzing existing methods, which can be categorized into pointwise or listwise approaches.

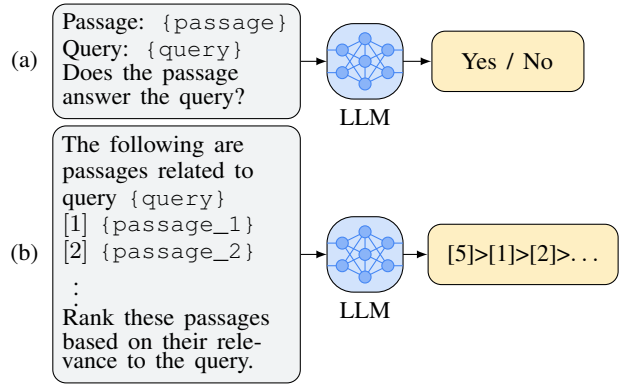
\begin{figure}
  \centering
  \input{figures/baselines}
  \caption{Two existing prompting methods for ranking: (a) the pointwise relevance generation approach and (b) the listwise permutation approach.}  
  \label{figure:baselines}
\end{figure}

\subsection{Pointwise approaches}
\label{sec:pointwise}

Pointwise approaches are the major methods prior to very recent listwise approaches discussed in Section~\ref{sec:listwise}. There are two popular methods, relevance generation~\citep{liang2022holistic} and query generation~\citep{sachan2022improving, drozdov2023parade}. Figure~\ref{figure:baselines} (a) shows the prompt used for relevance generation. The relevance score $s_i$ is defined as:
\begin{equation}
s_i =
\begin{cases}
1 + p(\textrm{Yes}), \textrm{if output Yes}\\
1 - p(\textrm{No}), \textrm{if output No}    
\end{cases}
\end{equation}
where $p(\textrm{Yes})$ and $p(\textrm{No})$ denote the probabilities of LLMs generating `Yes' and `No' respectively. Meanwhile query generation approach asks LLMs to generate a query based on the document ("Please write a question based on this passage. Passage: \{\{passage\}\} Question:"), and measures the probability of generating the actual query. Readers can refer to~\citet{sachan2022improving} for more details.

There are two major issues with pointwise approaches. First, pointwise relevance prediction requires the model to output \emph{calibrated} pointwise predictions so that they can be used for comparisons in sorting. This is not only very difficult to achieve across prompts~\citep{desai2020calibration}, but also unnecessary for ranking, which only requires \textit{relative} ordering, a major focus of the learning to rank field~\citep{8186875}. Also, pointwise methods will not work for generation API, which is common, such as GPT-4, since it requires the log probability of the desired predictions to perform sorting.

\subsection{Listwise approaches}
\label{sec:listwise}

Very recently, two parallel works~\citep{baidugpt,jimmygpt} explore listwise approaches, by directly inserting the query and a list of documents into a prompt. Both methods feed a partial list of 10 or 20 documents every time and perform a sliding window approach due to the prompt length constraints. Figure~\ref{figure:baselines} (b) shows a simplified version of the listwise ranking prompt. Both works explored text-davinci-003, i.e., InstructGPT~\citep{ouyang2022training} with 175B parameters, showing significantly worse performance than fine-tuned baseline rankers. ~\citet{baidugpt} were able to further explore gpt-3.5-turbo (the model behind ChatGPT) and GPT-4. Only the GPT-4 based approach could achieve competitive results, which is based on the blackbox, commercial, and giant (1T estimated parameters~\citep{WinNT,baktash2023gpt}) system, without academic publication discussing technical details (\citet{openai2023gpt4} mainly focused on evaluations). 

The issues are again due to the difficulty of the listwise ranking task for LLMs. ~\citet{baidugpt} show that there are frequent prediction failures with the following patterns: 
\begin{itemize}[noitemsep,topsep=0pt,parsep=0pt,partopsep=0pt]
    \item Missing: When LLMs only outputs a partial list of the input documents. 
    \item Rejection: LLMs refuse to perform the ranking task and produce irrelevant outputs. 
    \item Repetition: LLMs output the same document more than once.
    \item Inconsistency: The same list of documents have different output rankings when they are fed in with different order or context. 
\end{itemize}
 In fact, we tried the same prompt from~\citep{baidugpt} on the FLAN-UL2 model with 20B parameters, and found very few of the outputs to be usable. The model will either just output few documents (e.g., "[1]"), an ordered list based on id (e.g. "[3] > [2] > [1] ..."), or text which is not parseable.
 
Different from pointwise approaches, listwise approaches can only use the generation API -- getting the log probability of all listwise permutations is prohibitively expensive. In other words, there is no easy solution if the generation API does not output desired results, which is common. These methods will fall back to the initial ranking, and due to the high failure rate, the results are highly sensitive to input ordering.
 
These observations are not entirely surprising. Existing popular LLMs are generally not specifically pre-trained or fine-tuned against ranking tasks. However, we show that LLMs do have a sense of pairwise relative comparisons, which is much simpler than requiring a calibrated pointwise relevance estimation or outputting a permutation for a list of documents.

\section{Pairwise ranking prompting}
\label{sec:pairwise}
We propose Pairwise Ranking Prompting (PRP) for ranking with LLMs. We describe the basic pairwise prompting unit, how it supports both generation and scoring APIs, and propose several variants of PRP with different ranking strategies and efficiency properties.

\subsection{Prompting design}
Our pairwise ranking prompt is simple and intuitive, as shown in Figure~\ref{figure:prompting}. The exact prompt template is shown in Appendix F. This pairwise prompting will serve the basic computation unit in all PRP variants, which we denote as $u(q, d_1, d_2)$ for a query $q$ and two documents $d_1$ and $d_2$.

\begin{figure}
  \centering
  \input{figures/prompting}
  \caption{An illustration of pairwise ranking prompting. The scores in scoring mode represent the log-likelihood of the model generating the target text given the prompt. See the exact prompt template in Appendix~\ref{sec:app:reproduce}}
  \label{figure:prompting}
\end{figure}
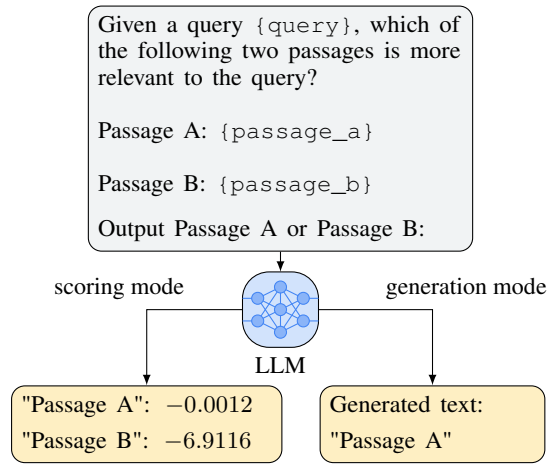

PRP naturally supports both generation API and scoring API. The latter is made possible since we only have two expected outputs ("Passage A" and "Passage B") for LLM inquiries. Since using scoring mode can mitigate potential issues when the generation API generates irrelevant outputs, our main results are based on the scoring mode, though we show there are very few prediction failures and provide comparisons between these two modes in Section~\ref{sec:ablation}.

Since it is known that LLMs can be sensitive to text orders in the prompt~\citep{lu2022fantastically, liu2023lost}, for each pair of documents, we will inquire the LLM twice by swapping their order: $u(q, d_1, d_2)$ and $u(q, d_2, d_1)$. Such simple de-biasing method is difficult for listwise methods due to their combinatorial nature.

The output of the pairwise ranking prompting is a local ordering of $d_1 > d_2$ or $d_2 > d_1$ if both promptings make consistent decisions, and $d_1 = d_2$ otherwise. Next we discuss three variants of PRP using the output of pairwise ranking prompting as the computation unit. We note that pairwise comparison can serve as the basic computation unit of many algorithms (e.g., selection algorithm) and leave other alternatives for future work.

\subsection{All pair comparisons}
We enumerate all pairs and perform a global aggregation to generate a score $s_i$ for each document $d_i$. We call this approach PRP-Allpair. Specifically, we have:  
\begin{equation}
\label{eq:allpair}
    s_i = 1 \cdot \sum_{j \neq i}\mathbb{I}_{d_i > d_j} + 0.5 \cdot \sum_{j \neq i}\mathbb{I}_{d_i = d_j}.
\end{equation}
Intuitively, if the LLM consistently prefers $d_i$ over another document $d_j$, $d_i$ gets one point. When LLM is not sure by producing conflicting or irrelevant results (for the generation API), each document gets half a point. There might be ties for the aggregated scores, in which case we fall back to initial ranking. In this work, we use \eqref{eq:allpair} which works for both scoring and generation APIs, and note there could be other ways to weight the scoring function, such as leveraging prediction probabilities in scoring mode.

PRP-Allpair favors simple implementation (all LLM API calls can be executed in parallel), and is highly insensitive to input ordering. It essentially ranks documents with win ratio, which has strong theoretical guarantees~\citep{shah2018simple}. The clear drawback is its costly $O(N^2)$ calls to LLM APIs, where $N$ is the number of documents to be ranked for each query.

\subsection{Sorting-based}
We note that efficient sorting algorithms, such as Quicksort and Heapsort, depend on pairwise comparisons. We can use the pairwise preferences from LLMs as the comparator for sorting algorithms. We use Heapsort in this paper due to its guaranteed $O(N \log N)$ computation complexity. We call this approach PRP-Sorting.

PRP-Sorting favors lower computation complexity than PRP-Allpair while also being large insensitive to input orders. Even though pairwise comparisons are not guaranteed to be transitive, we show robust empirical performance in the experiments, and leave applying methods with theoretical guarantees~\citep{ailon2008aggregating,bai2023sorting} for future work.

\subsection{Sliding window}

\begin{figure}
  \centering
  \input{figures/sliding_window}
  \caption{An illustration of one pass of our sliding window approach. Starting from right to left, we compare each document pair and swap it if the LLM output disagrees with the initial ranking. $K$ such passes will ensure a high-performing top-$K$ ranking.}
  \label{figure:sliding_window}
\end{figure}
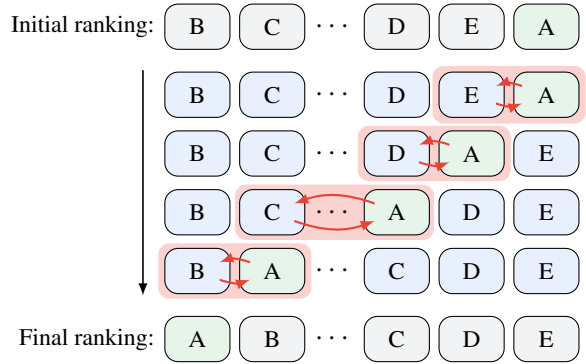

We introduce a sliding window approach that is able to further bring down the computation complexity. One sliding window pass is similar to one pass in the Bubble Sort algorithm: Given an initial ranking, we start from the bottom of the list, compare and swap document pairs with a stride of 1 on-the-fly based on LLM outputs. One pass only requires $O(N)$ time complexity. See Figure~\ref{figure:sliding_window} for an illustration.

By noticing that ranking usually only cares about Top-$K$ ranking metrics, we can perform $K$ passes, where $K$ is small, even if thousands of documents are ranked~\citep{zhuang2022rankt5}. We call this approach PRP-Sliding-K. 

PRP-Sliding-K has favorable time complexity but may have high dependency on input order. In experiments we show surprisingly good results with PRP-Sliding-10, without being very sensitive to input ordering empirically in Section~\ref{sec:ablation}).

\begin{table*}[h]
\small
\centering
\caption{Comparison of pointwise, listwise, and pairwise approaches. $N$ is the number of documents to be ranked for each query. $O(N)$ for listwise approach is based on sliding window since other options are not practical. See discussion on "Require Calibration" in Section~\ref{sec:pointwise}.}
\begin{tabular}{c|c|c|c|c}
\toprule
Method & \# of LLM API Calls & Generation API & Scoring API & Require Calibration \\
\midrule
Pointwise & $O(N)$ & No & Yes & Yes \\
Listwise & $O(N)$ & Yes & No & No \\
Pairwise & $O(N^2), O(N\log N), O(N)$ & Yes & Yes & No\\
\bottomrule
\end{tabular}
\label{tbl:comparison}
\end{table*}

\subsection{Remarks}
In this work, we focus on open-sourced LLMs that are easily accessible to academic researchers, and do not require inquiry of commercial LLM APIs, alleviating some monetary constraints. Also, the LLMs do not need to be finetuned in the prompting-based setting. We briefly summarize the properties of pointwise, pairwise, and listwise ranking promptings in Table~\ref{tbl:comparison}, showing pairwise ranking prompting has several favorable properties.

\section{Experiments on TREC DL datasets}

\subsection{Datasets and Metrics}
TREC is a widely used benchmark dataset in information retrieval research. We use the test sets of
the 2019 and 2020 competitions: TREC-DL2019 and TREC-DL2020, which provide dense human relevance annotations for each of their 43 and 54 queries. Both use the MS MARCO v1 passage corpus, which contains 8.8 million passages. All comparisons are based on the reranking of top 100 passages retrieved by BM25~\citep{Lin_etal_SIGIR2021_Pyserini} for each query. This is the same setting as existing work~\citep{baidugpt,jimmygpt}.

\subsection{Methods}
\label{sec:methods1}
We evaluate PRP variants based on open-sourced LLMs, including FLAN-T5-XL, FLAN-T5-XXL~\citep{flant5}, and FLAN-UL2~\citep{ul2}, which have significantly smaller model sizes (3B, 11B, 20B) than alternatives, and are easily accessible to academic researchers. We report PRP variants including PRP-Allpair, PRP-Sorting, and PRP-Sliding-K.

We consider the following supervised baselines, all trained on the in-domain MS MARCO dataset:
\begin{itemize}[noitemsep,topsep=0pt,parsep=0pt,partopsep=0pt]
    \item monoBERT~\citep{nogueira2019passage}: A cross-encoder re-ranker based on BERT-large.
    \item monoT5~\citep{monot5}: A sequence-to-sequence re-ranker that uses T5 to calculate the relevance score with pointwise ranking loss.
    \item RankT5~\citep{zhuang2022rankt5}: A re-ranker that uses T5 and listwise ranking loss.
\end{itemize}

We also consider the following unsupervised LLM-based baselines:
\begin{itemize}[noitemsep,topsep=0pt,parsep=0pt,partopsep=0pt]
    \item Unsupervied Passage Re-ranker (UPR)~\citep{sachan2022improving}: The \textit{pointwise} approach based on query generation, see Section~\ref{sec:pointwise}.
    \item Relevance Generation (RG)~\citep{liang2022holistic}: The \textit{pointwise} approach based on relevance generation, see Section~\ref{sec:pointwise}.
    \item RankGPT~\citep{baidugpt}: The \textit{listwise} prompting based approach using various GPT based LLMs. As discussed in Section~\ref{sec:listwise}, we tried the listwise prompt on FLAN-T5 and FLAN-UL2 models and the outputs are not usable, so we only report results with large blackbox LLMs.
    \item Listwise Reranker with a Large language model (LRL)~\citep{jimmygpt}: A similar approach to RankGPT with slightly different prompt design.
\end{itemize}

\subsection{Main Results}

\begin{table*}[t]
\large
\centering
\caption{Results on TREC-DL2019 and TREC-DL2020 datasets by reranking top 100 documents retrieved by BM25. Best overall model is in boldface, best and second best unsupervised LLM method are underlined and italicized respectively, for each metric. All unsupervised LLM methods use BM25 to resolve prediction conflicts or failures. *OpenAI has not publicly released the model parameters and the numbers are based on public estimates~\citep{WinNT,baktash2023gpt}}
\resizebox{\textwidth}{!}{
\begin{tabular}{l|l|l||ccc|ccc}
\toprule
\multirow{2}{*}{Method} & \multirow{2}{*}{LLM}& \multirow{2}{*}{Size} & \multicolumn{3}{c|}{\textbf{TREC-DL2019}} & \multicolumn{3}{c}{\textbf{TREC-DL2020}}\\
&&& NDCG@1 & NDCG@5 & NDCG@10 & NDCG@1 & NDCG@5 & NDCG@10 \\
\midrule
 BM25 & NA & NA & 54.26 & 52.78 & 50.58 & 57.72 & 50.67 & 47.96 \\
\midrule
\multicolumn{8}{c}{\textbf{Supervised Methods}} \\
\midrule
monoBERT & BERT & 340M &  79.07 & 73.25 &70.50 &78.70 &70.74 &67.28 \\
monoT5 & T5 & 220M &79.84 & 73.77 &71.48 &77.47& 69.40 &66.99\\
monoT5  & T5 & 3B &79.07 & 73.74  &71.83 &80.25 &72.32& 68.89\\
RankT5 & T5 & 3B & 79.07 & 75.66 & 72.95 & 80.86 & 73.05 & 69.63 \\
\midrule
\multicolumn{8}{c}{\textbf{Unsupervised LLM Methods}} \\
\midrule
 LRL & text-davinci-003 & 175B & - & - & 65.80 & - & - & 62.24 \\
 RankGPT & gpt-3 & 175B &   50.78 & 50.77 & 49.76 & 50.00 & 48.36 & 48.73 \\
 RankGPT & text-davinci-003 & 175B &  69.77 & 64.73 & 61.50 &  69.75 & 58.76 & 57.05 \\
 RankGPT & gpt-3.5-turbo & 154B* &   \textit{82.17} &71.15 &65.80 &79.32 &66.76& 62.91 \\
RankGPT & gpt-4  & 1T* &  \underline{\textbf{82.56}}& \underline{\textbf{79.16}} &\underline{\textbf{75.59}}& 78.40 &74.11 &\textit{70.56} \\
UPR & FLAN-T5-XXL & 11B & 62.79 & 62.07 & 62.00 & 64.20 & 62.05 & 60.34 \\
RG & FLAN-T5-XXL & 11B & 67.05 & 65.41 & 64.48 & 65.74 & 66.40 & 62.58  \\
UPR & FLAN-UL2 & 20B & 53.10 & 57.68 & 58.95 & 64.81 & 61.50 & 60.02 \\
RG & FLAN-UL2 & 20B & 70.93 & 66.81 & 64.61 & 75.62 & 66.85 & 65.39 \\
\midrule
PRP-Allpair & FLAN-T5-XL & 3B & 74.03 & 71.73 & 69.75 & 79.01 & 72.22 & 68.12 \\
PRP-Sorting & FLAN-T5-XL & 3B & 77.52 & 71.88 & 69.28 & 74.38 & 69.44 & 65.87 \\
PRP-Sliding-10 & FLAN-T5-XL & 3B & 75.58 & 71.23 & 68.66 & 75.62 & 69.00 & 66.59 \\
PRP-Allpair & FLAN-T5-XXL & 11B & 72.09 & 71.28& 69.87& 82.41 & 74.16 & 69.85 \\
PRP-Sorting & FLAN-T5-XXL & 11B & 74.42 & 69.62 & 67.81 & 72.53 & 71.28 & 67.77 \\
PRP-Sliding-10 & FLAN-T5-XXL & 11B & 64.73 & 69.49 & 67.00 & 75.00 & 70.76 & 67.35 \\
PRP-Allpair & FLAN-UL2 & 20B & 73.64 & 74.77& 72.42 &\textit{85.19} & \textit{74.73} & \underline{\textbf{70.68}}\\
PRP-Sorting & FLAN-UL2 & 20B & 74.42 & 73.60 & 71.88 & 84.57 & 72.52 & 69.43 \\
PRP-Sliding-10 & FLAN-UL2 & 20B & 78.29 & \textit{75.49} & \textit{72.65} & \underline{\textbf{85.80}} & \underline{\textbf{75.35}} & 70.46 \\

\bottomrule
\end{tabular}
}
\label{tbl:result1}
\end{table*}

Our main results are shown in Table~\ref{tbl:result1}. Overall we are able to achieve very encouraging results using PRP. We have the following observations:
\begin{itemize}[noitemsep,topsep=0pt,parsep=0pt,partopsep=0pt]
    \item PRP variants based on FLAN-UL2 with 20B parameters can achieve best results on all metrics on TREC-DL2020, and are only second to the blackbox, commercial gpt-4 based solution on NDCG@5 and NDCG@10 on TREC-DL2019, which has an estimated 50X larger model size. Our best methods outperform RankGPT based on text-davinci-003 with 175B parameters by over 10\% on all ranking metrics, and are competitive to supervised methods on all ranking metrics.
    \item Results on FLAN-T5-XL and FLAN-T5-XXL are also competitive, showing that PRP generalizes to smaller LLMs due to the significant simplicity of the pairwise ranking comparisons. They generally work even better than the gpt-3.5.turbo based solution (10X - 50X in size) on the more stable NDCG@5 and NDCG@10 metrics, and outperforms text-davinci-003 based solution on all ranking metrics.
    \item It is encouraging to see good results from efficient PRP variants. For example, the sliding window variants generally get very robust ranking performance and we get some of the best metrics from this variant. This observation alleviates some efficiency concerns of pairwise ranking approaches. 
\end{itemize}

\section{Experiments on BEIR datasets}
\subsection{Datasets and metrics}
BEIR~\citep{thakur2021beir} consists of diverse retrieval tasks and domains. Following ~\citep{baidugpt} we choose the test sets of Covid, Touche, DBPedia, SciFact, Signal, News, and Robust04. Following the convention of related research, we report NDCG@10 for each dataset and the average NDCG@10.

\subsection{Methods}
We use the \textit{same} prompt template from TREC datasets for all BEIR datasets, which is consistent for all compared unsupervised LLM-based baselines. This is in contrast to methods such as ~\citep{dai2022promptagator} that require prior knowledge to design different prompts for different datasets, which may be difficult in practice and will lead to unfair comparisons. 

For supervised methods, in addition to the baselines in Section~\ref{sec:methods1}, we add TART ~\citep{tart}, a supervised instruction-tuned passage re-ranker trained on 37 datasets, including over 5 million instances. The model is initialized from FLAN-T5-XL.

For unsupervised LLM methods, we also report RG and UPR as in Section~\ref{sec:methods1}. We include RankGPT with gpt-3.5-turbo. We do not include the GPT-4 numbers reported in ~\citep{baidugpt}, which used GPT-4 to \textit{rerank} top results from gpt-3.5-turbo due to the significant cost. It essentially performed an ensemble of two re-ranking models, which is unfair and impractical. We also do not include LRL since it was not evaluated on the BEIR collection. See more discussions of baselines in Appendix~\ref{sec:app:datasets}.

\subsection{Main Results}
\begin{table*}[t]
\large
\centering
\caption{Results (NDCG@10) on BEIR datasets. All models re-rank the same BM25 top-100 passages.
Best overall model is in boldface, best and second best unsupervised LLM method are underlined and italicized respectively, for each metric. All unsupervised LLM methods use BM25 to resolve prediction conflicts or failures.}
\resizebox{\textwidth}{!}{
\begin{tabular}{l|l|l||cccccccc}
\toprule
Method & LLM & Size & Covid &Touche& DBPedia &SciFact &Signal &News &Robust04& Avg\\
\midrule
 BM25 & NA & NA & 59.47 &\textbf{44.22} &31.80 &67.89& 33.05 &39.52 &40.70 &45.23 \\
\midrule
\multicolumn{8}{c}{\textbf{Supervised Methods}} \\
\midrule
monoBERT & BERT & 340M &  70.01 &31.75 &41.87 &71.36 &31.44& 44.62 &49.35 &48.63 \\
monoT5 & T5 & 220M &78.34 &30.82 &42.42 &73.40 &31.67 &46.83 &51.72 &50.74\\
monoT5  & T5 & 3B &80.71 &32.41 &44.45 &\textbf{76.57} &32.55 &48.49 &56.71 &53.13\\
RankT5 & T5 & 3B & 82.00 & 37.62 & 44.19 &76.86& 31.80& 48.15& 52.76& 53.34\\
TART-Rerank & T5 & 3B & 75.10 &27.46& 42.53& 74.84& 25.84& 40.01& 50.75 &48.08\\
\midrule
\multicolumn{8}{c}{\textbf{Unsupervised LLM Methods}} \\
\midrule
UPR & FLAN-T5-XXL & 11B & 72.64 & 21.56 & 35.14 &73.54 &30.81 & 42.99 & 47.85 & 46.36\\
RG & FLAN-T5-XXL & 11B & 70.31 & 22.10& 31.32 &63.43 & 26.89 & 37.34 &51.56 & 43.28\\
UPR & FLAN-UL2 & 20B & 70.69 & 23.68 & 34.64 &71.09 & 30.33 & 41.78 & 47.52 & 45.68\\
RG & FLAN-UL2 & 20B & 70.22 & 24.67& 30.56 &64.74 & 29.68&43.78 &53.00 & 45.24\\
RankGPT & gpt-3.5-turbo &154B& 76.67  &36.18 &44.47 &70.43& 32.12 &\textit{48.85} &50.62 &51.33 \\
\midrule
PRP-Allpair & FLAN-T5-XL & 3B & 81.86 &   26.93& 44.63 & 73.25&32.08 & 46.52 & 54.02& 51.33\\
PRP-Sorting & FLAN-T5-XL & 3B & 80.41 &   28.23& 42.84& 67.94&30.95 & 42.95 & 50.07& 49.06 \\
PRP-Sliding-10 & FLAN-T5-XL & 3B & 77.58   &\textit{40.48}& 44.77& 73.43&\underline{\textbf{35.62}}& 46.45& 50.74& 52.72 \\
PRP-Allpair & FLAN-T5-XXL & 11B & 79.62 &   29.81& 41.41& \textit{74.23} &32.22& 47.68& \underline{\textbf{56.76}}& 51.67 \\
PRP-Sorting & FLAN-T5-XXL & 11B& 78.75 &   29.61& 39.23& 70.10 &31.28& 44.68& 53.01& 49.52 \\
PRP-Sliding-10 & FLAN-T5-XXL & 11B& 74.39 &   \underline{41.60}&42.19 & 72.46&35.12& 47.26& 52.38& 52.20 \\
PRP-Allpair & FLAN-UL2 & 20B & \underline{\textbf{82.30}} & 29.71&\textit{45.94}& \underline{75.70} & 32.26 &48.04 &\textit{55.49}& \textit{52.78} \\
PRP-Sorting & FLAN-UL2 & 20B &  \textit{82.29} & 25.80&44.53 & 67.07& 32.04 &45.37 &51.45& 49.79 \\
PRP-Sliding-10 & FLAN-UL2 & 20B &  79.45 & 37.89& \underline{\textbf{46.47}}& 73.33 & \textit{35.20} &\underline{\textbf{49.11}} &53.43& \underline{\textbf{53.55}} \\

\bottomrule
\end{tabular}
}
\label{tbl:beir_main}
\end{table*}

The main results are shown in Table~\ref{tbl:beir_main}. Overall we are able to achieve encouraging results using PRP, validating its robustness across different domains. We have the following observations:
\begin{itemize}[noitemsep,topsep=0pt,parsep=0pt,partopsep=0pt]
   \item PRP variants based on FLAN-UL2 with 20B parameters can achieve best overall results on the collection. 
   \item  PRP variants generate the best ranking metrics on all datasets among unsupervised LLM methods. PRP outperforms the blackbox commercial RankGPT solution by 4.2\%, and pointwise LLM-based solutions by over 10\% in general. Noticably, PRP-Sliding-10 with FLAN-UL2 outperforms RankGPT on \textit{all} 7 datasets, showing its strong generalization. 
   \item PRP performs favorably with supervised methods. PRP-Sliding-10 with FLAN-UL2 can slightly outperform the state-of-the-art RankT5 ranker on average, and outperform RankT5 on 5 out of 7 datasets.  
    \item Results on FLAN-T5-XL and FLAN-T5-XXL are again competitive, some variants can even outperform RankGPT.
\end{itemize}

\section{Ablation studies}
\label{sec:ablation}
We perform several ablative studies to gain a deeper understanding of the PRP framework in terms of its robustness and generality. 

\paragraph{Robustness to input ordering.} We show the robustness of PRP to input ordering. One issue of listwise ranking prompting approaches is their sensitivity to input ordering. This is because the ranking will fall back to the initial order when LLM prediction fails, which is very common for the difficult listwise formulation. In Table~\ref{tbl:inverse} we show results of different methods by inverting the initial order from BM25.

\begin{table*}[h]
\small
\centering
\caption{Input order sensitivity results on the TREC-DL2019 dataset.}
\begin{tabular}{c|c|c||c|c|c}
\toprule
Method & LLM & Init Order & NDCG@1 & NDCG@5 & NDCG@10  \\
\midrule
RankGPT & gpt-3.5-turbo & BM25 & 82.17 & 71.15 & 65.80 \\
RankGPT & gpt-3.5-turbo & Inverse BM25 & 36.43 & 31.79 & 32.77  \\
\midrule
PRP-Allpair & FLAN-UL2-20B & BM25& 73.64 & 74.77 & 72.42  \\
PRP-Allpair & FLAN-UL2-20B & Inverse BM25& 74.42 & 74.48 & 72.40  \\
\midrule
PRP-Sliding-1 & FLAN-UL2-20B & BM25& 78.29 & 62.15 & 57.58 \\
PRP-Sliding-1 & FLAN-UL2-20B & Inverse BM25 & 71.32 & 32.72 & 26.04 \\
\midrule
PRP-Sliding-10 & FLAN-UL2-20B & BM25& 78.29 & 75.49 & 72.65 \\
PRP-Sliding-10 & FLAN-UL2-20B & Inverse BM25 & 71.32  & 67.91 & 64.84 \\
\bottomrule
\end{tabular}
\label{tbl:inverse}
\end{table*}

As expected, PRP-Allpair is quite robust to initial ordering, and PRP-Sliding-1 will suffer for metrics other than NDCG@1. PRP-Sliding-10 is quite robust since it focuses on Top-K ranking metrics.

\paragraph{Comparison of scoring mode and generation mode.} Our results above are all based on the scoring mode, since PRP only need to get scores for two candidate outputs ("Passage A" and "Passage B") and it is easy to get probabilities from open-sourced LLMs. Here we compare against PRP performance using scoring vs generation mode in Table~\ref{tbl:mode}, which will shed light on how PRP works on generation-only LLM APIs.

\begin{table*}[h]
\large
\centering
\caption{Results on TREC-DL2019 and TREC-DL2020 datasets using scoring vs generation mode for PRP.}
\vspace{-0.4em}
\resizebox{\textwidth}{!}{
\begin{tabular}{l|l|l||ccc|ccc}
\toprule
\multirow{2}{*}{Method} & \multirow{2}{*}{LLM}& \multirow{2}{*}{Mode} & \multicolumn{3}{c|}{\textbf{TREC-DL2019}} & \multicolumn{3}{c}{\textbf{TREC-DL2020}}\\
&&& NDCG@1 & NDCG@5 & NDCG@10 & NDCG@1 & NDCG@5 & NDCG@10 \\
\midrule
PRP-Allpair & FLAN-T5-XL & Scoring & 74.03 & 71.73 & 69.75 & 79.01 & 72.22 & 68.12 \\
PRP-Allpair & FLAN-T5-XL & Generation  &74.03 & 71.68 & 69.59 & 79.01 & 71.54 & 67.75    \\
PRP-Allpair & FLAN-T5-XXL & Scoring & 72.09 & 71.28& 69.87& 82.41 & 74.16 & 69.85 \\
PRP-Allpair & FLAN-T5-XXL & Generation &72.09 &71.61 & 69.94&80.56&73.69&69.53 \\
PRP-Allpair & FLAN-UL2 & Scoring & 73.64 & 74.77& 72.42 &85.19 & 74.73 & 70.68\\
PRP-Allpair & FLAN-UL2 & Generation & 73.64 &74.84 &72.37& 85.19 & 74.74 & 70.69\\
\bottomrule
\end{tabular}
}
\label{tbl:mode}
\end{table*}

We can see that PRP is extremely robust to scoring vs generation API, even for smaller LLMs, showing its applicability to different LLMs systems. The results are intuitive - LLMs make few generation mistakes due to the simplicity of PRP. We found that there are only about 0.02\% predictions that do not follow the desired format, which is neglectable and in stark contrast to the the listwise approaches. 

\paragraph{Study on sliding window.} We further provide more study on the sliding window approach in Appendix~\ref{sec:sliding}, including different number of passes and the performance of forward (instead of backward) pass.

\section{Discussion}
\paragraph{Extendability.} The design of PRP in this paper biases towards simplicity and generality. For example, we decribe the algorithm and report results based on generation API, so PRP is applicable to both commercial black-box LLMs and open-sourced white-box LLMs. The performance may further improve via more sophisticated prompt design, and leveraging extra information such as the score values from the scoring API, which is usually available for white-box LLMs. We provide some results of PRP on a commercial LLMs in Appendix~\ref{sec:ulm24b} where performance can be further improved.

\paragraph{Reproducibility.} We used the same prompt template for all 9 datasets evaluated in the paper, showing the generality and power of pairwise ranking prompting in text ranking. As we focus on open-sourced LLMs, and only use standard aggregation methods (win counting, sorting, and sliding window), our experimental results are easy to reproduce. Still, we plan to release pairwise inference results on all 9 datasets and the 3 open-source LLMs to facilitate future research. In specific, we will release the data in json format, which includes query/document information for each pair (including ids, text, label, retrieval rank and scores), together with the actual prompt, the generated text, and its score. The specific prompt template and a data sample can be found at Appendix~\ref{sec:app:reproduce}

\paragraph{Cost and Efficiency.} We discussed different efficient variants of PRP. Also, our results are based on LLMs that are easily approachable for academic researchers~\citep{alpaca}, alleviating the need to call commercial APIs. However, further reducing the number of calls to LLMs is still an interesting research direction, such as leveraging active learning techniques. The distillation of LLM rankers to servable models in large-scale systems is also an important future direction~\citep{sun2023instruction, qin2023rd}. 

\paragraph{Data Leakage from LLMs.} We note there is minimal label leakage issues as we leverage open-sourced LLMs with clear documentations, while it is not clear for blackbox commercial LLMs. The comparisons with existing pointwise and listwise approaches on the same LLMs are also fair. Please see a more comprehensive examination on data leakage in Appendix~\ref{sec:discuss}.

\section{Related Work}
We did a detailed review and analysis of the most relevant existing efforts for ranking with LLMs, including pointwise and listwise approaches in Section~\ref{sec:analysis}. These works and ours focus on the challenging unsupervised text ranking setting with LLMs without providing any demonstrations, conducting any fine-tuning, or training of an additional model. Prior to the recent efforts on ranking with LLMs, most work focus on the supervised 
learning to rank problem~\citep{8186875,dasalc} by fine-tuning Pre-trained Language Models (PLMs) such as T5~\citep{monot5,zhuang2022rankt5} or BERT~\citep{nogueira2019passage, zhuang2021ensemble}, which serve as very strong baselines. Very recently some work fine-tunes LLMs or distills from black-box LLMs~\citep{pradeep2023rankvicuna}, which is different from our setting.

There has been a strong recent interest in exploring information retrieval in general with LLMs based approaches~\citep{zhu2023large}, due to the importance of the applications and the power of LLMs to understand textual queries and documents~\citep{dai2022promptagator,tay2022transformer,wang2023query2doc, jagerman2023query,bonifacio2022inpars}. Several works leverage the generation power of LLMs to generate training data to train an additional downstream retrieval or ranking model, typically in the few-shot setting~\citep{dai2022promptagator}, which is a very different setting from ours. Recent methods in this family of methods such as Inpars~\citep{bonifacio2022inpars} still significantly underperforms fine-tuned baselines. ExaRanker~\citep{ferraretto2023exaranker} uses LLMs to generate explanations for ranking decisions, and uses such explanations in ranking model fine-tuning, showing limited ranking performance benefits (the major benefit was on data efficiency). HyDE~\citep{gao2022precise} uses LLMs to augment queries by generating hypothetical documents for unsupervised retrieval. These works do not directly explore the retrieval or ranking capability of LLMs, but mainly use LLMs as auxiliary tools to complement traditional paradigms, possibly limiting the benefits that LLMs can provide. New paradigms such as Differentiable Search Index (DSI)~\citep{tay2022transformer, wang2022neural} directly use Transformer memory to index documents for retrieval. 

Using pairwise comparisons with LLMs is a general paradigm, such as reward modeling using pairwise preferences~\citep{christiano2017deep, rafailov2024direct, liu2024lipo}. LLMs are used as evaluators to compare generative outputs (such as text summary)~\citep{liu2023calibrating, liusie2024llm}. SC~\citep{yan2023basis} performs structured comparative reasoning to predict text preferences in various applications. 1SL~\citep{macavaney2023one} estimates relevance with reference to an anchor positive query-document pair \textit{per query}, even for the test set, so the setting may not be practical and is very different from our standard text ranking setting. A concurrent work~\citep{10.1145/3604915.3610646} studied pairwise prompting in recommender systems, which is a substantially different application and their method still largely fall behind state-of-the-art models with sufficient data. 
The novelty of our work lies in leveraging the general and simple pairwise prompting paradigm to the important text ranking task, granting LLMs capabilities that no prior work can, by performing competitively with state-of-the-art fine-tuned models and methods that only work with giant blackbox LLMs.
\section{Conclusion}
In this paper, we propose to use pairwise prompting with LLMs for text ranking tasks. To the best of our knowledge, these are the first published results demonstrating very competitive ranking performance using moderate-sized, open-sourced LLMs. The key insights are the observation of the difficulties of LLMs handling ranking tasks in the existing pointwise and listwise formulations. Our proposed Pairwise Ranking Prompting (PRP) is effective in reducing the burden of LLMs and shows robust performance on 9 datasets. We also discuss efficiency concerns and ways to mitigate them, and several benefits of PRP, such as insensitivity to input ordering and support for both generation and scoring LLM APIs.

\section{Limitations}
We do not use GPT models (though we compare with them using results from other papers) in this work due to various constraints and the focus on open-sourced LLMs. Testing the performance of our methods on such models is meaningful benchmarking effort. Also, this work mainly focused on empirical ranking results, while more theoretically grounded methods exist, such as those for sorting from noisy comparisons~\citep{bai2023sorting}, which may be explored in the future. Last but not least, we discuss the potential data leakage issue (for all LLM-based methods) in Appendix~\ref{sec:discuss}.

\bibliography{custom}

\appendix
\onecolumn

\section{More results on PRP-Sliding-K}
\label{sec:sliding}
We show more results on PRP-Sliding-K variants to better understand the behaviors, including multiple backward passes and a forward pass variant\footnote{Backward pass indicates starting from the bottom result with the lowest BM25 score, and vice versa.}. The results are shown in Table~\ref{tbl:sliding19} and Table~\ref{tbl:sliding20} on TREC-DL2019 and TREC-DL2020 with consistent behaviors.

\begin{table*}[h]
\small
\centering
\caption{Sliding window results on the TREC-DL2019 dataset.}
\begin{tabular}{c|c|c||c|c|c}
\toprule
Method & LLM & Strategy & NDCG@1 & NDCG@5 & NDCG@10  \\
\midrule
PRP-Sliding & FLAN-UL2-20B & 1 Forward & 63.95 & 57.31 & 54.10 \\
PRP-Sliding & FLAN-UL2-20B & 1 Backward & 78.29 & 62.15 & 57.58\\
PRP-Sliding & FLAN-UL2-20B & 2 Backward & 78.29 & 67.01 & 61.52\\
PRP-Sliding & FLAN-UL2-20B & 3 Backward & 78.29 & 70.72 & 64.60\\
PRP-Sliding & FLAN-UL2-20B & 10 Backward & 78.29 & 75.49 & 72.65\\

\bottomrule
\end{tabular}
\label{tbl:sliding19}
\end{table*}

\begin{table*}[h]
\small
\centering
\caption{Sliding window results on the TREC-DL2020 dataset.}
\begin{tabular}{c|c|c||c|c|c}
\toprule
Method & LLM & Strategy & NDCG@1 & NDCG@5 & NDCG@10 \\
\midrule
PRP-Sliding & FLAN-UL2-20B & 1 Forward & 65.74 & 54.72 & 51.21 \\
PRP-Sliding & FLAN-UL2-20B & 1 Backward & 85.80 & 61.60 &  57.06\\
PRP-Sliding & FLAN-UL2-20B & 2 Backward & 85.80 & 66.51&  61.11\\
PRP-Sliding & FLAN-UL2-20B & 3 Backward & 85.80 & 71.06&  63.45\\
PRP-Sliding & FLAN-UL2-20B & 10 Backward & 85.80 & 75.35&  70.46\\
\bottomrule
\end{tabular}
\label{tbl:sliding20}
\end{table*}

The results are easy to interpret:
\begin{itemize}[noitemsep,topsep=0pt,parsep=0pt,partopsep=0pt]
    \item The behavior is similar to BubbleSort: Strong NDCG@1 can already be achieved with one backward pass. As we conduct more passes, other Top-K ranking metrics get better. 
    \item Forward pass does not work well, which is intuitive, since it mainly performs demotion and is much less efficient in bringing good results to the top.   
\end{itemize}

\section{Result of PRP on commercial LLMs}
\label{sec:ulm24b}
Though the focus on the work is to show the power of PRP on moderate-sized LLMs, we further perform evaluation on two datasets with a black-box commercial LLM, text-bison, from Google (https://cloud.google.com/vertex-ai/docs/generative-ai/model-reference/text), which should be comparable to gpt-3.5-turbo. The results can be further improved when compared with our main results on open-sourced LLMs, showing the generality of PRP. Further evaluation on more powerful LLMs such as gpt-4 is meaningful future work. 
\begin{table}[h]
\small
\centering
\caption{Results on PRP-Allpair with the text-bison model on TREC-DL2019 and TREC-DL2020.}
\begin{tabular}{c|c|c|c|}
\toprule
Method & LLM & NDCG@10 DL19 & NDCG@10 DL20  \\
\midrule
RankGPT & gpt-3.5-turbo & 65.80 & 62.91\\
RankGPT & gpt-4 & 75.59 & 70.56 \\
PRP-Allpair & FLAN-UL2-20B & 72.42 & 70.68 \\ 
PRP-Allpair & text-bison & 73.81 & 71.66\\
\bottomrule
\end{tabular}
\label{tbl:ulm24b}
\end{table}

\section{More discussion on limitations and future work}
\label{sec:discuss}

\paragraph{Domain adaptation.} The datasets used in this paper are for the standard and important relevance-based text ranking. How LLMs can be adapted to non-standard ranking datasets, such as counter arguments in the ArguAna dataset~\citep{wachsmuth:2018a}, need more investigation. Our work can facilitate such explorations by providing approachable baselines.  

\paragraph{Data leakage.} We mainly use open-sourced FLAN models~\citep{wei2021finetuned} with clear documentations, which neither observed ranking supervision from any of the datasets we evaluated upon, nor was instruction fine-tuned on any ranking tasks. Also, the labels in the datasets are \textit{dense} human annotations for each query against many documents, which are not used in the open-sourced LLMs and are very different from the potential usage of document corpus during pre-training. These are in contrast to methods based blackbox LLMs such as ChatGPT or GPT-4~\citep{baidugpt} where the tuning details are unclear. We do note that FLAN models have a question answering task based on MSMARCO, which is not ranking specific, and is different from TREC-DL datasets in terms of queries and annotations, and is different from BEIR collection in all aspects. On the other hand, whether blackbox LLMs directly use TREC-DL datasets or BEIR datasets is unclear. Furthermore, the comparisons between different methods using the same LLM are fair - PRP always outperforms pointwise baselines by a large margin, and listwise prompting almost always fails on moderate LLMs. Avoiding data leakage in the era of LLM is generally challenging and more rigorous protocols may be needed. In this work, we avoided to use phrases such as ``zero-shot'' to try to avoid over-claims.

\section{More discussion on baseline and dataset selection}
\label{sec:app:datasets}
For the BEIR evaluation, we choose not to include the Promptagator++ ranker ~\citep{dai2022promptagator} since 1) It uses different prompts and fine-tuned models for each task, different from all other LLM methods. 2) The method was evaluated on a different set of BEIR tasks. Even for the shared tasks, it reranks top 200 results from a stronger retriever than BM25 so the numbers are not comparable. Nevertheless, zero-shot Promptagator++ performed significantly \textit{worse} than the monoT5 baseline in the paper (to be fair, the paper's focus was mainly on few-shot scenarios), while PRP compares favorably with monoT5. 

The only dataset we did not include, but \citep{baidugpt} included, from the BEIR collection, is the NFCorpus dataset. This is because the metrics using BM25 reported in \citep{baidugpt} on NFCorpus does not match ours and the public consensus numbers (while the numbers match for all selected datasets), so we exclude NFCorpus to avoid unfair comparisons possibly due to errors during their evaluation.

\section{Reproducibility}
\label{sec:app:reproduce}

\subsection{Pairwise Ranking Prompting Template}
We note that we used the \textbf{same prompt template for all 9 datasets} evaluated in the paper, showing the generality and power of pairwise ranking prompting in text ranking. Below is the prompt template:

\begin{tcolorbox}
Given a query \code{\{query\}}, which of the following two passages is more relevant to the query? \\

Passage A: \code{\{document$_1$\}} \\

Passage B: \code{\{document$_2$\}} \\

Output Passage A or Passage B:
\end{tcolorbox}

\subsection{Code and Data Release}
As we focus on open-sourced LLMs, and only use standard aggregation methods (win counting, sorting, and sliding window), our experimental results are easy to reproduce. We plan to release pairwise inference results on all 9 datasets and the 3 open-source LLMs to facilitate future research. In specific, we will release the data in the following json format, which includes query/document information for each pair (including ids, text, label, retrieval rank and scores), together with the actual prompt, the generated text, and its score. Below is an example on the Trec-DL2020 dataset with Flan-UL2:

\newpage
\begin{tcolorbox}
"document\_pair": [\{"document\_id": "8512412", "retriever\_rank": "50", "retriever\_score": "8.984600", "document": "When in Doubt, Take a Cab. Taxis might be expensive in Puerto Rico, but they are safe and available. At night, it's definitely the best way to get around. Look for the white taxis with the distinctive garita, or sentry box, icon painted on them.They are usually found at designated taxi stands.hen in Doubt, Take a Cab. Taxis might be expensive in Puerto Rico, but they are safe and available. At night, it's definitely the best way to get around. Look for the white taxis with the distinctive garita, or sentry box, icon painted on them.", "relevance": -1\}, \{"document\_id": "6623205", "retriever\_rank": "66", "retriever\_score": "8.812100", "document": "Thankfully, there are a couple of ways to prevent your whites from turning yellow: 1  Never bleach white clothing that is polyester or a polyester/cotton blend. 2  The chemical reaction between the bleach and the polyester almost always yields a yellowed result. 3  Consider a water softener if you have well-water.hankfully, there are a couple of ways to prevent your whites from turning yellow: 1  Never bleach white clothing that is polyester or a polyester/cotton blend. 2  Consider a water softener if you have well-water. 3  Minimize your use of bleach altogether.", "relevance": 1.0\}], \\

"query\_id": "1108651", \\

"query": "what the best way to get clothes white", \\

"prompt": "Given a query ``what the best way to get clothes white'', which of the following two passages is more relevant to the query? \\

Passage A: When in Doubt, Take a Cab. Taxis might be expensive in Puerto Rico, but they are safe and available. At night, it's definitely the best way to get around. Look for the white taxis with the distinctive garita, or sentry box, icon painted on them.They are usually found at designated taxi stands.hen in Doubt, Take a Cab. Taxis might be expensive in Puerto Rico, but they are safe and available. At night, it's definitely the best way to get around. Look for the white taxis with the distinctive garita, or sentry box, icon painted on them. \\

Passage B: Thankfully, there are a couple of ways to prevent your whites from turning yellow: 1  Never bleach white clothing that is polyester or a polyester/cotton blend. 2  The chemical reaction between the bleach and the polyester almost always yields a yellowed result. 3  Consider a water softener if you have well-water.hankfully, there are a couple of ways to prevent your whites from turning yellow: 1  Never bleach white clothing that is polyester or a polyester/cotton blend. 2  Consider a water softener if you have well-water. 3  Minimize your use of bleach altogether. \\

Output Passage A or Passage B:", \\

"generated\_text": "Passage B", \\

"prediction\_score": -0.0025123630184680223 \\

\end{tcolorbox}

\end{document}

%% file: math_commands.tex
\usepackage{amsmath,amsfonts,bm}

\def\eqref#1{equation~\ref{#1}}

\def\1{\bm{1}}

\DeclareMathAlphabet{\mathsfit}{\encodingdefault}{\sfdefault}{m}{sl}
\SetMathAlphabet{\mathsfit}{bold}{\encodingdefault}{\sfdefault}{bx}{n}



%% file: colors.tex
\definecolor{g900grey}{HTML}{202124}
\definecolor{g900blue}{HTML}{174EA6}
\definecolor{g900red}{HTML}{A50E0E}
\definecolor{g900yellow}{HTML}{E37400}
\definecolor{g900green}{HTML}{0D652D}
\definecolor{g700grey}{HTML}{5F6368}
\definecolor{g700blue}{HTML}{1967d2}
\definecolor{g700red}{HTML}{c5221f}
\definecolor{g700yellow}{HTML}{f29900}
\definecolor{g700green}{HTML}{188038}
\definecolor{g500grey}{HTML}{9AA0A6}
\definecolor{g500blue}{HTML}{4285F4}
\definecolor{g500red}{HTML}{EA4335}
\definecolor{g500yellow}{HTML}{FBBC04}
\definecolor{g500green}{HTML}{34A853}
\definecolor{g400blue}{HTML}{669DF6}
\definecolor{g300green}{HTML}{81C995}
\definecolor{g300grey}{HTML}{DADCE0}
\definecolor{g300blue}{HTML}{8AB4F8}
\definecolor{g200green}{HTML}{A8DAB5}
\definecolor{g200blue}{HTML}{AECBFA}
\definecolor{g100grey}{HTML}{F1F3F4}
\definecolor{g100green}{HTML}{CEEAD6}
\definecolor{g100blue}{HTML}{D2E3FC}
\definecolor{g050green}{HTML}{E6f4EA}
\definecolor{g050blue}{HTML}{E8F0FE}
\definecolor{g100red}{HTML}{FAD2CF}
\definecolor{g100yellow}{HTML}{FEEFC3}
\definecolor{g100green}{HTML}{CEEAD6}

%% file: figures/util.tex
\tikzset{
    neuralnetwork/.pic = {
        \node[draw, draw=black, minimum height=1.0cm, minimum width=1.0cm, text width=0.5cm, align=center, fill=g100blue, rounded corners=10pt] (llm_box) at (0, 0) {};
        
        \path[draw=g500blue] (-0.5, 0.150) -- (-0.30,  0.150);
        \path[draw=g500blue] (-0.5,-0.150) -- (-0.30, -0.150);
        
        \path[draw=g500blue] (-0.30, 0.150) -- (0,  0.30);
        \path[draw=g500blue] (-0.30, 0.150) -- (0,  0   );
        \path[draw=g500blue] (-0.30, 0.150) -- (0, -0.30);
        \path[draw=g500blue] (-0.30,-0.150) -- (0,  0.30);
        \path[draw=g500blue] (-0.30,-0.150) -- (0,  0   );
        \path[draw=g500blue] (-0.30,-0.150) -- (0, -0.30);
        
        \path[draw=g500blue] (0,-0.30) -- (0, 0   );
        \path[draw=g500blue] (0, 0   ) -- (0, 0.30);
        
        \path[draw=g500blue] (0, 0.30) -- (0.30,  0.150);
        \path[draw=g500blue] (0, 0.30) -- (0.30, -0.150);
        \path[draw=g500blue] (0, 0   ) -- (0.30,  0.150);
        \path[draw=g500blue] (0, 0   ) -- (0.30, -0.150);
        \path[draw=g500blue] (0,-0.30) -- (0.30,  0.150);
        \path[draw=g500blue] (0,-0.30) -- (0.30, -0.150);
        
        \path[draw=g500blue] (0.30, 0.150) -- (0.5, 0.150);
        \path[draw=g500blue] (0.30,-0.150) -- (0.5,-0.150);
        
        \draw[draw=g500blue,fill=g300blue](-0.30,  0.150) circle (2.25 pt);
        \draw[draw=g500blue,fill=g300blue](-0.30, -0.150) circle (2.25 pt);
        \draw[draw=g500blue,fill=g300blue]( 0   ,  0.30 ) circle (2.25 pt);
        \draw[draw=g500blue,fill=g300blue]( 0   ,  0    ) circle (2.25 pt);
        \draw[draw=g500blue,fill=g300blue]( 0   , -0.30 ) circle (2.25 pt);
        \draw[draw=g500blue,fill=g300blue]( 0.30,  0.150) circle (2.25 pt);
        \draw[draw=g500blue,fill=g300blue]( 0.30, -0.150) circle (2.25 pt);
    }
}

%% file: figures/baselines.tex
\begin{tikzpicture}[scale=0.8]

\node (prompt_pointwise_label) at (-5.6, 2) {\small (a)};
\node[draw, minimum height=1.5cm, minimum width=3.2cm, text width=3cm, align=left, rounded corners=5pt, fill=g100grey, execute at begin node=\setlength{\baselineskip}{1pt}]
  (prompt_pointwise) at (-3.1, 2)
  {{\small Passage: \texttt{\{passage\}} \\
    Query: \texttt{\{query\}} \\
    Does the passage answer the query?
   }};
\pic [local bounding box=llm_pointwise] at (0, 2) {neuralnetwork};
\node[below,align=center] at (llm_pointwise.south) {{\small LLM}};
\node[draw, minimum height=0.8cm, minimum width=2cm, text width=1.5cm, align=center, rounded corners=5pt, fill=g100yellow]
  (answer_pointwise) at (2.35, 2)
  {{\small Yes / No}};

\node (prompt_listwise_label) at (-5.6, -1.2) {\small (b)};
\node[draw, minimum height=1.5cm, minimum width=3.2cm, text width=3cm, align=left, rounded corners=5pt, fill=g100grey, execute at begin node=\setlength{\baselineskip}{1pt}]
  (prompt_listwise) at (-3.1, -1.2)
  {{\small The following are passages related to query \texttt{\{query\}}\\
  $[1]$ \texttt{\{passage\_1\}} \\
  $[2]$ \texttt{\{passage\_2\}} \\
  \;\,$\vdots$ \\
  Rank these passages based on their relevance to the query.
   }};
\pic [local bounding box=llm_listwise] at (0, -1.2) {neuralnetwork};
\node[below,align=center] at (llm_listwise.south) {{\small LLM}};
\node[draw, minimum height=0.8cm, minimum width=2.5cm, text width=2cm, align=center, rounded corners=5pt, fill=g100yellow]
  (answer_listwise) at (2.6, -1.2)
  {{\small [5]>[1]>[2]>$\ldots$}};
  
\path[->,>=latex,draw] (prompt_pointwise) -- (llm_pointwise);
\path[->,>=latex,draw] (llm_pointwise) -- (answer_pointwise);
\path[->,>=latex,draw] (prompt_listwise) -- (llm_listwise);
\path[->,>=latex,draw] (llm_listwise) -- (answer_listwise);

\end{tikzpicture}

%% file: figures/prompting.tex
\begin{tikzpicture}[scale=0.8]

\node[draw, minimum height=1.5cm, minimum width=5cm, text width=4.8cm, align=left, rounded corners=5pt, fill=g100grey, execute at begin node=\setlength{\baselineskip}{1pt}]
  (prompt) at (0, 0)
  {{\small Given a query \texttt{\{query\}}, which of the following two passages is more relevant to the query?\\\,\\Passage A: \texttt{\{passage\_a\}}\\\,\\Passage B: \texttt{\{passage\_b\}}\\\,\\Output Passage A or Passage B:}};

\pic [local bounding box=llm] at (0, -3.0) {neuralnetwork};
\node[below,align=center] at (llm.south) {{\small LLM}};

\node[draw, minimum height=0.7cm, minimum width=2.8cm, text width=2.7cm, align=left, fill=g100yellow, rounded corners=5pt]
  (generated) at (2.5, -4.9)
  {{\small Generated text: "Passage A"}};

\node[draw, minimum height=0.7cm, minimum width=3.4cm, text width=3.3cm, align=left, fill=g100yellow, rounded corners=5pt]
  (scoring) at (-2.2, -4.9)
  {{\small "Passage A": $-0.0012$\\"Passage B": $-6.9116$}};

\path[->,>=latex,draw]
    (prompt) -- (llm);

\path[->,>=latex,draw]
    (llm) -| node[near start,above right] {{\small generation mode}} (generated);

\path[->,>=latex,draw]
    (llm) -| node[near start,above left] {{\small scoring mode}} (scoring);

\end{tikzpicture}

%% file: figures/sliding_window.tex
\begin{tikzpicture}[scale=0.82]

\begin{scope}[shift={(0, 0)},local bounding box=toprow]
\node[draw, draw=black, minimum height=0.6cm, minimum width=0.8cm, text width=0.6cm, align=center, fill=g100grey, rounded corners=5pt] (passage_b) at (0  , 0) {\small B};
\node[draw, draw=black, minimum height=0.6cm, minimum width=0.8cm, text width=0.6cm, align=center, fill=g100grey, rounded corners=5pt] (passage_c) at (1.2, 0) {\small C};
\node (dots) at (2.2, 0) {$\cdots$};
\node[draw, draw=black, minimum height=0.6cm, minimum width=0.8cm, text width=0.6cm, align=center, fill=g100grey, rounded corners=5pt] (passage_d) at (3.2, 0) {\small D};
\node[draw, draw=black, minimum height=0.6cm, minimum width=0.8cm, text width=0.6cm, align=center, fill=g100grey, rounded corners=5pt] (passage_e) at (4.4, 0) {\small E};
\node[draw, draw=black, minimum height=0.6cm, minimum width=0.8cm, text width=0.6cm, align=center, fill=g050green, rounded corners=5pt] (passage_a) at (5.6, 0) {\small A};
\end{scope}
\node (init) [at=(toprow.west),align=right,anchor=east] {{\small Initial ranking:}};

\begin{scope}[shift={(0, -1.1)}]
\node[draw, draw=g100red, fill=g100red,minimum height=0.75cm, minimum width=2.0cm, rounded corners=4pt] (bg) at (5.0, 0) {};
\node[draw, draw=black, minimum height=0.6cm, minimum width=0.8cm, text width=0.6cm, align=center, fill=g050blue, rounded corners=5pt] (passage_b) at (0  , 0) {\small B};
\node[draw, draw=black, minimum height=0.6cm, minimum width=0.8cm, text width=0.6cm, align=center, fill=g050blue, rounded corners=5pt] (passage_c) at (1.2, 0) {\small C};
\node (dots) at (2.2, 0) {$\cdots$};
\node[draw, draw=black, minimum height=0.6cm, minimum width=0.8cm, text width=0.6cm, align=center, fill=g050blue, rounded corners=5pt] (passage_d) at (3.2, 0) {\small D};
\node[draw, draw=black, minimum height=0.6cm, minimum width=0.8cm, text width=0.6cm, align=center, fill=g050blue, rounded corners=5pt] (passage_e) at (4.4, 0) {\small E};
\node[draw, draw=black, minimum height=0.6cm, minimum width=0.8cm, text width=0.6cm, align=center, fill=g050green, rounded corners=5pt] (passage_a) at (5.6, 0) {\small A};
\draw [->,>=latex,g500red,line width=0.8pt] ([yshift=-4pt,xshift=-4pt]passage_e.east) to [out=-20,in=-160] ([yshift=-4pt,xshift=4pt]passage_a.west);
\draw [->,>=latex,g500red,line width=0.8pt] ([yshift=4pt,xshift=4pt]passage_a.west) to [out=160,in=20] ([yshift=4pt,xshift=-4pt]passage_e.east);
\end{scope}

\begin{scope}[shift={(0, -2.05)}]
\node[draw, draw=g100red, fill=g100red,minimum height=0.72cm, minimum width=2.0cm, rounded corners=4pt] (bg) at (3.8, 0) {};
\node[draw, draw=black, minimum height=0.6cm, minimum width=0.8cm, text width=0.6cm, align=center, fill=g050blue, rounded corners=5pt] (passage_b) at (0  , 0) {\small B};
\node[draw, draw=black, minimum height=0.6cm, minimum width=0.8cm, text width=0.6cm, align=center, fill=g050blue, rounded corners=5pt] (passage_c) at (1.2, 0) {\small C};
\node (dots) at (2.2, 0) {$\cdots$};
\node[draw, draw=black, minimum height=0.6cm, minimum width=0.8cm, text width=0.6cm, align=center, fill=g050blue, rounded corners=5pt] (passage_d) at (3.2, 0) {\small D};
\node[draw, draw=black, minimum height=0.6cm, minimum width=0.8cm, text width=0.6cm, align=center, fill=g050green, rounded corners=5pt] (passage_a) at (4.4, 0) {\small A};
\draw [->,>=latex,g500red,line width=0.8pt] ([yshift=-4pt,xshift=-5pt]passage_d.east) to [out=-20,in=-160] ([yshift=-4pt,xshift=5pt]passage_a.west);
\draw [->,>=latex,g500red,line width=0.8pt] ([yshift=4pt,xshift=5pt]passage_a.west) to [out=160,in=20] ([yshift=4pt,xshift=-5pt]passage_d.east);
\node[draw, draw=black, minimum height=0.6cm, minimum width=0.8cm, text width=0.6cm, align=center, fill=g050blue, rounded corners=5pt] (passage_e) at (5.6, 0) {\small E};
\end{scope}

\begin{scope}[shift={(0, -3.0)}]
\node[draw, draw=g100red, fill=g100red,minimum height=0.72cm, minimum width=2.6cm, rounded corners=4pt] (bg) at (2.2, 0) {};
\node[draw, draw=black, minimum height=0.6cm, minimum width=0.8cm, text width=0.6cm, align=center, fill=g050blue, rounded corners=5pt] (passage_b) at (0  , 0) {\small B};
\node[draw, draw=black, minimum height=0.6cm, minimum width=0.8cm, text width=0.6cm, align=center, fill=g050blue, rounded corners=5pt] (passage_c) at (1.2, 0) {\small C};
\node (dots) at (2.2, 0) {$\cdots$};
\node[draw, draw=black, minimum height=0.6cm, minimum width=0.8cm, text width=0.6cm, align=center, fill=g050green, rounded corners=5pt] (passage_a) at (3.2, 0) {\small A};
\draw [->,>=latex,g500red,line width=0.8pt] ([yshift=-4pt,xshift=-5pt]passage_c.east) to [out=-20,in=-160] ([yshift=-4pt,xshift=5pt]passage_a.west);
\draw [->,>=latex,g500red,line width=0.8pt] ([yshift=4pt,xshift=5pt]passage_a.west) to [out=160,in=20] ([yshift=4pt,xshift=-5pt]passage_c.east);
\node[draw, draw=black, minimum height=0.6cm, minimum width=0.8cm, text width=0.6cm, align=center, fill=g050blue, rounded corners=5pt] (passage_d) at (4.4, 0) {\small D};
\node[draw, draw=black, minimum height=0.6cm, minimum width=0.8cm, text width=0.6cm, align=center, fill=g050blue, rounded corners=5pt] (passage_e) at (5.6, 0) {\small E};
\end{scope}

\begin{scope}[shift={(0, -3.95)}]
\node[draw, draw=g100red, fill=g100red,minimum height=0.72cm, minimum width=2.0cm, rounded corners=4pt] (bg) at (0.6, 0) {};
\node[draw, draw=black, minimum height=0.6cm, minimum width=0.8cm, text width=0.6cm, align=center, fill=g050blue, rounded corners=5pt] (passage_b) at (0  , 0) {\small B};
\node[draw, draw=black, minimum height=0.6cm, minimum width=0.8cm, text width=0.6cm, align=center, fill=g050green, rounded corners=5pt] (passage_a) at (1.2, 0) {\small A};
\draw [->,>=latex,g500red,line width=0.8pt] ([yshift=-4pt,xshift=-5pt]passage_b.east) to [out=-20,in=-160] ([yshift=-4pt,xshift=5pt]passage_a.west);
\draw [->,>=latex,g500red,line width=0.8pt] ([yshift=4pt,xshift=5pt]passage_a.west) to [out=160,in=20] ([yshift=4pt,xshift=-5pt]passage_b.east);
\node (dots) at (2.2, 0) {$\cdots$};
\node[draw, draw=black, minimum height=0.6cm, minimum width=0.8cm, text width=0.6cm, align=center, fill=g050blue, rounded corners=5pt] (passage_c) at (3.2, 0) {\small C};
\node[draw, draw=black, minimum height=0.6cm, minimum width=0.8cm, text width=0.6cm, align=center, fill=g050blue, rounded corners=5pt] (passage_d) at (4.4, 0) {\small D};
\node[draw, draw=black, minimum height=0.6cm, minimum width=0.8cm, text width=0.6cm, align=center, fill=g050blue, rounded corners=5pt] (passage_e) at (5.6, 0) {\small E};
\end{scope}

\begin{scope}[shift={(0, -5.05)},local bounding box=finalrow]
\node[draw, draw=black, minimum height=0.6cm, minimum width=0.8cm, text width=0.6cm, align=center, fill=g050green, rounded corners=5pt] (passage_a) at (0  , 0) {\small A};
\node[draw, draw=black, minimum height=0.6cm, minimum width=0.8cm, text width=0.6cm, align=center, fill=g100grey, rounded corners=5pt] (passage_b) at (1.2, 0) {\small B};
\node (dots) at (2.2, 0) {$\cdots$};
\node[draw, draw=black, minimum height=0.6cm, minimum width=0.8cm, text width=0.6cm, align=center, fill=g100grey, rounded corners=5pt] (passage_c) at (3.2, 0) {\small C};
\node[draw, draw=black, minimum height=0.6cm, minimum width=0.8cm, text width=0.6cm, align=center, fill=g100grey, rounded corners=5pt] (passage_d) at (4.4, 0) {\small D};
\node[draw, draw=black, minimum height=0.6cm, minimum width=0.8cm, text width=0.6cm, align=center, fill=g100grey, rounded corners=5pt] (passage_e) at (5.6, 0) {\small E};
\end{scope}
\node (final) [at=(finalrow.west),align=right,anchor=east] {{\small Final ranking:}};
\path[->,>=latex,draw,line width=0.6pt] ([xshift=-10pt,yshift=-20pt]init.east) -- ([xshift=-10pt,yshift=20pt]final.east);

\end{tikzpicture}